\documentclass[aps,preprint,nofootinbib,floatfix]{revtex4}%
\usepackage{amssymb}
\usepackage{amsmath}
\usepackage{epsfig}
\usepackage{amsfonts}
\usepackage{graphicx}%
\providecommand{\U}[1]{\protect\rule{.1in}{.1in}}
\begin{document}
\title{Exponential fall-off Behavior of Regge Scatterings in Compactified Open String Theory}
\author{Song He}
\email{hesong@ihep.ac.cn}
\affiliation{Institute of High Energy Physics, Chinese Academy of Sciences, Beijing 100039,
China }
\affiliation{Kavli Institute for Theoretical Physics China, CAS, Beijing 100190, China}
\author{Jen-Chi Lee}
\email{jcclee@cc.nctu.edu.tw}
\affiliation{Department of Electrophysics, National Chiao-Tung University and Physics
Division, National Center for Theoretical Sciences, Hsinchu, Taiwan, R.O.C.}
\affiliation{Kavli Institute for Theoretical Physics China, CAS, Beijing 100190, China}
\author{Yi Yang}
\email{yiyang@mail.nctu.edu.tw}
\affiliation{Department of Electrophysics, National Chiao-Tung University and Physics
Division, National Center for Theoretical Sciences, Hsinchu, Taiwan, R.O.C.}
\affiliation{Kavli Institute for Theoretical Physics China, CAS, Beijing 100190, China}

\begin{abstract}
We calculate massive string scattering amplitudes of compactified open string
in the Regge regime. We extract the \textit{complete} infinite ratios among
high-energy amplitudes of different string states in the fixed angle regime
from these Regge string scattering amplitudes. The complete ratios calculated
by this indirect method include and extend the subset of ratios calculated
previously \cite{Compact,Compact2} by the more difficult direct fixed angle
calculation. In this calculation of \textit{compactified} open string
scattering, we discover a realization of arbitrary real values $L$ in the
identity Eq.(\ref{14}), rather than integer value only in all previous
high-energy string scattering amplitude calculations. The identity in
Eq.(\ref{14}) was explicitly proved recently in \cite{LYAM} to link fixed
angle and Regge string scattering amplitudes. In addition, we discover a
kinematic regime with stringy highly winding modes, which shows the unusual
exponential fall-off behavior in the Regge string scattering. This is in
complementary with a kenematic regime discovered previously \cite{Compact2},
which shows the unusual power-law behavior in the high-energy fixed angle
compactified string scatterings.


\end{abstract}
\maketitle

\bigskip%
\setcounter{equation}{0}
\renewcommand{\theequation}{\arabic{section}.\arabic{equation}}%

\section{Introduction}

There are three fundamental characteristics of high-energy fixed angle string
scattering amplitudes \cite{GM, Gross, GrossManes}, which are not shared by
the field theory scattering. These are the softer exponential fall-off
behavior (in contrast to the hard power-law behavior of field theory
scatterings), the infinite Regge-pole structure of the form factor and the
existence of infinite number of linear relations \cite{ChanLee1,ChanLee2,
CHL,CHLTY,PRL,paperB,susy,Closed,Dscatt,Decay}, or stringy symmetries,
discovered recently among high-energy string scattering amplitudes of
different string states. An important new ingredient to derive these linear
relations is the zero-norm states (ZNS) \cite{ZNS1,ZNS3,ZNS2} in the old
covariant first quantized (OCFQ) string spectrum, in particular, the
identification of inter-particle symmetries induced by the inter-particle ZNS
\cite{ZNS1} in the spectrum. Other approaches related to this development can
be found in \cite{MooreWest}.

Recently, following an old suggestion of Mende \cite{Mende}, two of the
present authors \cite{Compact} calculate high-energy fixed angle massive
scattering amplitudes of closed bosonic string with some coordinates
compactified on the torus. The calculation was extended to the compactified
open string scatterings \cite{Compact2}. An infinite number of linear
relations among high-energy scattering amplitudes of different string states
were obtained in the fixed angle or Gross kinematic regime (GR). The UV
behavior in the GR shows the usual soft exponential fall-off behavior. These
results are reminiscent of the existence of an infinite number of massive ZNS
in the compactified closed \cite{Lee} and open \cite{Lee2} string spectrums
constructed previously. In addition, it was discovered that, for some
kinematic regime with super-highly winding modes at \textit{fixed angle}, the
so-called Mende kinematic regime (MR), these infinite linear relations break
down and, simultaneously, the string amplitudes enhance to hard power-law
behavior at high energies instead of the usual soft exponential fall-off behavior.

In this paper, we calculate high-energy \textit{small angle} or Regge string
scattering amplitudes \cite{RR1,RR2,RR3,RR4,RR5,RR6,DL,KP} of open bosonic
string with one coordinates compactified on the torus. The results can be
generalized to more compactified coordinates. It is shown that there is no
linear relations among Regge scattering amplitudes as expected. However, as in
the case of noncompactified Regge string scattering amplitude calculation
\cite{bosonic,bosonic2,RRsusy}, we can deduce the infinite GR ratios in the
fixed angle from these compactified Regge string scattering amplitudes. We
stress that the GR ratios calculated in the present paper by this indirect
method from the Regge calculation are for the most general high-energy vertex
rather than only a subset of GR ratios obtained directly from the fixed angle
calculation \cite{Compact,Compact2}. In this calculation, we have used a set
of master identities Eq.(\ref{14}) to extract the GR ratios from Regge
scattering amplitudes. Mathematically, the complete proof of these identities
for \textit{arbitrary real values} $L$ was recently worked out in \cite{LYAM}
by using an identity of signless Stirling number of the first kind in
combinatorial theory. The proof of the identity for $L=0,1,$ was previously
given in \cite{bosonic,bosonic2,RRsusy} based on a set of identities of signed
Stirling number of the first kind \cite{MK}. It is interesting to see that,
physically, the identities for arbitrary real values $L$ can only be realized
in high-energy compactified string scatterings considered in this paper. All
other high-energy string scatterings calculated previously
\cite{bosonic,bosonic2,RRsusy} correspond to integer values of $L$ only. A
recent work on string D-particle scatterings \cite{LMY} also gave integer
values $L$.

More importantly, we discover an exponential fall-off behavior of high-energy
compactified open string scatterings in a kinematic regime with highly winding
modes at small angle. The existence of this regime was conjectured in
\cite{Compact2}. However, no Regge scatterings were calculated there and thus
the results for the small angle scatterings extracted from the fixed angle
calculation were not completed and fully reliable \cite{bosonic,bosonic2}. The
discovery of the soft exponential fall-off behavior in this kinematic regime
with small angle in compactified string scatterings is in complementary with a
kenematic regime discovered previously \cite{Compact,Compact2}, which shows
the unusual power-law behavior in the high-energy fixed angle compactified
string scatterings. This paper is organized as following. In section II, we
set up the kinematics. In section III, we review the fixed angle compactified
string scatterings. Section IV is devoted to the compactified Regge string
scatterings. We first calculate the Regge string scattering amplitudes and
extract the most general fixed angle ratios from these Regge amplitudes. We
then derive a Regge regime which shows an unusual exponential fall-off
behavior. A brief conclusion is made in section V.%

\setcounter{equation}{0}
\renewcommand{\theequation}{\arabic{section}.\arabic{equation}}%

\section{Kinematics Set-up}

We consider 26D open bosonic string with one coordinate compactified on
$S^{1}$ with radius $R$. It is straightforward to generalize our calculation
to more compactified coordinates. The mode expansion of the compactified
coordinate is%
\begin{equation}
X^{25}\left(  \sigma,\tau\right)  =x^{25}+K^{25}\tau+i\sum_{k\neq0}%
\frac{\alpha_{k}^{25}}{k}e^{-ik\tau}\cos n\sigma
\end{equation}
where $K^{25}$ is the canonical momentum in the $X^{25}$ direction%
\begin{equation}
K^{25}=\frac{2\pi J-\theta_{l}+\theta_{i}}{2\pi R}.
\end{equation}
Note that $J$ is the quantized momentum and we have included a nontrivial
Wilson line with $U(n)$ Chan-Paton factors, $i,l=1,2...n.$, which will be
important in the later discussion. The mass spectrum of the theory is
\begin{equation}
M^{2}=\left(  K^{25}\right)  ^{2}+2\left(  N-1\right)  \equiv\left(
\frac{2\pi J-\theta_{l}+\theta_{i}}{2\pi R}\right)  ^{2}+\hat{M}^{2}
\label{mass}%
\end{equation}
where we have defined level mass as $\hat{M}^{2}=2\left(  N-1\right)  $ and
$N=\sum_{k\neq0}\alpha_{-k}^{25}\alpha_{k}^{25}+\alpha_{-k}^{\mu}\alpha
_{k}^{\mu},\mu=0,1,2...24.$ We are going to consider 4-point correlation
function in this paper. In the center of momentum frame, the kinematic can be
set up to be \cite{Compact,Compact2}%

\begin{align}
k_{1}  &  =\left(  +\sqrt{p^{2}+M_{1}^{2}},-p,0,-K_{1}^{25}\right)  ,\\
k_{2}  &  =\left(  +\sqrt{p^{2}+M_{2}^{2}},+p,0,+K_{2}^{25}\right)  ,\\
k_{3}  &  =\left(  -\sqrt{q^{2}+M_{3}^{2}},-q\cos\phi,-q\sin\phi,-K_{3}%
^{25}\right)  ,\\
k_{4}  &  =\left(  -\sqrt{q^{2}+M_{4}^{2}},+q\cos\phi,+q\sin\phi,+K_{4}%
^{25}\right)
\end{align}
where $p$ is the incoming momentum, $q$ is the outgoing momentum and $\phi$ is
the center of momentum scattering angle. In the high-energy limit, one
includes only momenta on the scattering plane, and we have included the fourth
component for the compactified direction as the internal momentum. The
conservation of the fourth component of the momenta implies%
\begin{equation}
\sum_{m}K_{m}^{25}=\sum_{m}(\frac{2\pi J_{m}-\theta_{l,m}+\theta_{i,m}}{2\pi
R})=0.
\end{equation}
Note that%
\begin{equation}
k_{i}^{2}=(K_{i}^{25})^{2}-M_{i}^{2}=-\hat{M}_{i}^{2}. \label{yi}%
\end{equation}
We have%
\begin{align}
-k_{1}\cdot k_{2}  &  =\sqrt{p^{2}+M_{1}^{2}}\cdot\sqrt{p^{2}+M_{2}^{2}}%
+p^{2}+K_{1}^{25}K_{2}^{25}\label{k1k2}\\
&  =\dfrac{1}{2}\left(  s+k_{1}^{2}+k_{2}^{2}\right)  =\dfrac{1}{2}s-\frac
{1}{2}\left(  \hat{M}_{1}^{2}+\hat{M}_{2}^{2}\right)  ,\nonumber\\
-k_{2}\cdot k_{3}  &  =-\sqrt{p^{2}+M_{2}^{2}}\cdot\sqrt{q^{2}+M_{3}^{2}%
}+pq\cos\phi+K_{2}^{25}K_{3}^{25}\label{t}\\
&  =\dfrac{1}{2}\left(  t+k_{2}^{2}+k_{3}^{2}\right)  =\dfrac{1}{2}t-\frac
{1}{2}\left(  \hat{M}_{2}^{2}+\hat{M}_{3}^{2}\right)  ,\nonumber\\
-k_{1}\cdot k_{3}  &  =-\sqrt{p^{2}+M_{1}^{2}}\cdot\sqrt{q^{2}+M_{3}^{2}%
}-pq\cos\phi-K_{1}^{25}K_{3}^{25}\label{u}\\
&  =\dfrac{1}{2}\left(  u+k_{1}^{2}+k_{3}^{2}\right)  =\dfrac{1}{2}u-\frac
{1}{2}\left(  \hat{M}_{1}^{2}+\hat{M}_{3}^{2}\right) \nonumber
\end{align}
where $s,t$ and $u$ are the Mandelstam variables with%
\begin{equation}
s+t+u=\sum_{i}\hat{M}_{i}^{2}=2\left(  N-4\right)  .
\end{equation}
Note that the Mandelstam variables defined above are not the usual
$25$-dimensional Mandelstam variables in the scattering process since we have
included the internal momentum $K_{i}^{25}$ in the definition of $k_{i}$. In
order to define the Regge or fixed momentum transfer regime, we define the momenta%

\begin{align}
\widehat{k}_{1}  &  =\left(  +\sqrt{p^{2}+\hat{M}_{1}^{2}},-p,0,0\right)  ,\\
\widehat{k}_{2}  &  =\left(  +\sqrt{p^{2}+\hat{M}_{2}^{2}},+p,0,0\right)  ,\\
\widehat{k}_{3}  &  =\left(  -\sqrt{q^{2}+\hat{M}_{3}^{2}},-q\cos\phi
,-q\sin\phi,0\right)  ,\\
\widehat{k}_{4}  &  =\left(  -\sqrt{q^{2}+\hat{M}_{4}^{2}},+q\cos\phi
,+q\sin\phi,0\right)
\end{align}
and the corresponding $25$-dimensional Mandelstam variables%

\begin{align}
-\widehat{k}_{1}\cdot\widehat{k}_{2}  &  =\sqrt{p^{2}+\hat{M}_{1}^{2}}%
\cdot\sqrt{p^{2}+\hat{M}_{2}^{2}}+p^{2}=\dfrac{1}{2}\left(  s_{25}-\hat{M}%
_{1}^{2}-\hat{M}_{2}^{2}\right)  ,\label{new1}\\
-\widehat{k}_{2}\cdot\widehat{k}_{3}  &  =-\sqrt{p^{2}+\hat{M}_{2}^{2}}%
\cdot\sqrt{q^{2}+\hat{M}_{3}^{2}}+pq\cos\phi=\dfrac{1}{2}\left(  t_{25}%
-\hat{M}_{2}^{2}-\hat{M}_{3}^{2}\right)  ,\label{new2}\\
-\widehat{k}_{1}\cdot\widehat{k}_{3}  &  =-\sqrt{p^{2}+\hat{M}_{1}^{2}}%
\cdot\sqrt{q^{2}+\hat{M}_{3}^{2}}-pq\cos\phi=\dfrac{1}{2}\left(  u_{25}%
-\hat{M}_{1}^{2}-\hat{M}_{3}^{2}\right)  \label{new3}%
\end{align}
where%

\begin{equation}
s_{25}+t_{25}+u_{25}=\sum_{i}\hat{M}_{i}^{2}=2\left(  N-4\right)  .
\label{new4}%
\end{equation}
In the high-energy limit, we define the polarizations on the scattering plane
to be%

\begin{align}
e^{P}  &  =\frac{1}{M_{2}}\left(  \sqrt{p^{2}+M_{2}^{2}},p,0,0\right)  ,\\
e^{L}  &  =\frac{1}{M_{2}}\left(  p,\sqrt{p^{2}+M_{2}^{2}},0,0\right)  ,\\
e^{T}  &  =\left(  0,0,1,0\right)
\end{align}
where the fourth component refers to the compactified direction. The center of
mass energy $E$ is defined as (for large $p,q$)%
\begin{equation}
E=\dfrac{1}{2}\left(  \sqrt{p^{2}+M_{1}^{2}}+\sqrt{p^{2}+M_{2}^{2}}\right)
=\dfrac{1}{2}\left(  \sqrt{q^{2}+M_{3}^{2}}+\sqrt{q^{2}+M_{4}^{2}}\right)  .
\end{equation}
The projections of the momenta on the scattering plane can be calculated to be
(here we only list the ones we will need for our calculation)%
\begin{align}
e^{P}\cdot k_{1}  &  =-\frac{1}{M_{2}}\left(  \sqrt{p^{2}+M_{1}^{2}}%
\sqrt{p^{2}+M_{2}^{2}}+p^{2}\right) \label{KR1}\\
e^{L}\cdot k_{1}  &  =-\frac{p}{M_{2}}\left(  \sqrt{p^{2}+M_{1}^{2}}%
+\sqrt{p^{2}+M_{2}^{2}}\right) \label{KR2}\\
e^{T}\cdot k_{1}  &  =0 \label{KR3}%
\end{align}
and
\begin{align}
e^{P}\cdot k_{3}  &  =\frac{1}{M_{2}}\left(  \sqrt{q^{2}+M_{3}^{2}}\sqrt
{p^{2}+M_{2}^{2}}-pq\cos\phi\right)  ,\label{KR4}\\
e^{L}\cdot k_{3}  &  =\frac{1}{M_{2}}\left(  p\sqrt{q^{2}+M_{3}^{2}}%
-q\sqrt{p^{2}+M_{2}^{2}}\cos\phi\right)  ,\label{KR5}\\
e^{T}\cdot k_{3}  &  =-q\sin\phi. \label{KR6}%
\end{align}
%

\setcounter{equation}{0}
\renewcommand{\theequation}{\arabic{section}.\arabic{equation}}%

\section{Fixed Angle Regime}

We begin with a brief review of high energy string scatterings for the
noncompactified 26D open bosonic string in the GR. That is in the kinematic
regime $s,-t\rightarrow\infty$, $t/s\approx-\sin^{2}\frac{\theta}{2}$= fixed
(but $\theta\neq0$) where $s,t$ and $u$ are the Mandelstam variables for the
noncompactified momenta and $\theta$ is the 26D CM scattering angle. It was
shown \cite{CHLTY,PRL} that for the 26D open bosonic string the only states
that will survive the high-energy limit at mass level $M_{2}^{2}=2(N-1)$ are
of the form
\begin{equation}
\left\vert N,2m,r\right\rangle \equiv(\alpha_{-1}^{T})^{N-2m-2r}(\alpha
_{-1}^{L})^{2m}(\alpha_{-2}^{L})^{r}\left\vert 0,k_{2}\right\rangle .
\label{1}%
\end{equation}
It can be shown that the high-energy vertex in Eq.(\ref{1}) are conformal
invariants up to a subleading term in the high-energy expansion. Note that
$e^{P}$ approaches to $e^{L}$ in the GR, and the scattering plane is defined
by the spatial components of $e^{L}$ and $e^{T}$. Polarizations perpendicular
to the scattering plane are ignored because they are kinematically suppressed
for four point scatterings in the high-energy limit. One can then use the
saddle-point method to calculate the high energy scattering amplitudes. For
simplicity, we choose $k_{1}$, $k_{3}$ and $k_{4}$ to be tachyons and the
final result of the ratios of high energy, fixed angle string scattering
amplitude are \cite{CHLTY,PRL}%
\begin{equation}
\frac{T^{(N,2m,r)}}{T^{(N,0,0)}}=\left(  -\frac{1}{M_{2}}\right)
^{2m+r}\left(  \frac{1}{2}\right)  ^{m+r}(2m-1)!!. \label{2}%
\end{equation}

We now review the results obtained previously for the compactified open string
scatterings at fixed angle $\phi=$ finite \cite{Compact2}. For simplicity, the
second vertex was chosen to be%

\begin{equation}
\left\vert N,0,r,i,l\right\rangle =\left(  \alpha_{-1}^{T}\right)
^{N-2r}\left(  \alpha_{-2}^{L}\right)  ^{r}\left\vert k_{2},l_{2}%
,i,l\right\rangle \label{Nr}%
\end{equation}
at mass level $\hat{M}_{2}^{2}=2\left(  N-1\right)  ,$ which was scattered
with three "tachyon" states (with $\hat{M}_{1}^{2}=\hat{M}_{3}^{2}=\hat{M}%
_{4}^{2}=-2$). The high-energy fixed angle open string scattering amplitudes
with one compactified coordinate were calculated to be (the trace factor due
to Chan-Paton was ignored) \cite{Compact2}%
\begin{align}
T^{(N,0,r,i,l)} &  \simeq\left(  -iq\sin\phi\right)  ^{N}\left(
-\frac{\left(  p\sqrt{q^{2}+M_{3}^{2}}-q\sqrt{p^{2}+M_{2}^{2}}\cos\phi\right)
}{M_{2}q^{2}\sin^{2}\phi}\right)  ^{r}\nonumber\\
&  \cdot\sum_{j=0}^{r}\binom{r}{j}\left[  -\frac{p\left(  \sqrt{p^{2}%
+M_{1}^{2}}+\sqrt{p^{2}+M_{2}^{2}}\right)  }{\left(  p\sqrt{q^{2}+M_{3}^{2}%
}-q\sqrt{p^{2}+M_{2}^{2}}\cos\phi\right)  }\right]  ^{j}\nonumber\\
&  \cdot B\left(  -1-\dfrac{1}{2}s,-1-\frac{1}{2}t\right)  \left(
-1-\dfrac{1}{2}s\right)  _{N-2j}\left(  -1-\frac{1}{2}t\right)  _{2j}\left(
2+\dfrac{1}{2}u\right)  _{N}^{-1}\label{power}%
\end{align}
where $(a)_{j}=a(a+1)(a+2)...(a+j-1)$ is the Pochhammer symbol, and
$(a)_{j}=a^{j}$ for large $a$ and fixed $j.$

\subsection{Fixed Winding Modes}

In the Gross regime, $p^{2}\gg K_{i}^{2}$ and $p^{2}\gg N$, Eq.(\ref{power})
reduces to%
\begin{equation}
T^{(N,0,r,i,l)}\simeq\left(  -iE\frac{\sin\frac{\phi}{2}}{\cos\frac{\phi}{2}%
}\right)  ^{N}\left(  -\frac{1}{2M_{2}}\right)  ^{r}\cdot B\left(  -1-\frac
{1}{2}s,-1-\frac{1}{2}t\right)  . \label{beta}%
\end{equation}
For each fixed mass level $N$, we have the linear relation for the scattering
amplitudes%
\begin{equation}
\frac{T^{(N,0,r,i,l)}}{T^{(N,0,0,i,l)}}=\left(  -\frac{1}{2M_{2}}\right)  ^{r}
\label{linear}%
\end{equation}
with ratios consistent with our previous result in Eq.(\ref{2}). Note that in
Eq.(\ref{beta}) there is an exponential fall-off factor in the high-energy
expansion of the beta function. The infinite linear relation in
Eq.(\ref{linear}) "soften" the high-energy behavior of string scatterings in
the GR.

\subsection{Super-highly Winding Modes}

We next consider a more interesting regime, the Mende kinematic regime (MR)
\cite{Compact2}. For the case of $\phi=$ finite, the only choice to achieve UV
power-law behavior is to require (we choose $(K_{1}^{25})^{2}\simeq(K_{2}%
^{25})^{2}\simeq(K_{3}^{25})^{2}\simeq(K_{4}^{25})^{2}$)%
\begin{equation}
\left(  K_{i}^{25}\right)  ^{2}\gg p^{2}\gg N. \label{newpower}%
\end{equation}
In order to explicitly show that this choice of kinematic regime does lead to
UV power-law behavior, it was shown that in this regime
\begin{equation}
s=\text{ constant} \label{const}%
\end{equation}
in the open string scattering amplitudes. This in turn gives the desire
power-law behavior of high-energy compactified open string scattering in
Eq.(\ref{power}). On the other hand, it can be shown that the linear relations
break down as expected in this regime. For the choice of kinematic regime in
Eq.(\ref{newpower}) , Eq.(\ref{k1k2}) and Eq.(\ref{const}) imply
\begin{equation}
\lim_{p\rightarrow\infty}\frac{\sqrt{p^{2}+M_{1}^{2}}\cdot\sqrt{p^{2}%
+M_{2}^{2}}+p^{2}}{K_{1}^{25}K_{2}^{25}}=\lim_{p\rightarrow\infty}\frac
{\sqrt{p^{2}+M_{1}^{2}}\cdot\sqrt{p^{2}+M_{2}^{2}}+p^{2}}{\left(  \frac{2\pi
l_{1}-\theta_{j,1}+\theta_{i,1}}{2\pi R}\right)  \left(  \frac{2\pi
l_{2}-\theta_{j,2}+\theta_{i,2}}{2\pi R}\right)  }=-1. \label{rev}%
\end{equation}
For finite momenta $J_{1}$and $J_{2}$, the power-law behavior can be achieved
by scattering of string states with "super-highly" winding nontrivial Wilson
lines%
\begin{equation}
(\theta_{i,1}-\theta_{l,1})\rightarrow\infty\text{, \ }(\theta_{i,2}%
-\theta_{l,2})\rightarrow-\infty. \label{wilson}%
\end{equation}
Note that the directions of momenta $K_{1}^{25}$ and $K_{2}^{25}$ are
\textit{opposite}. Since $\left(  K_{i}^{25}\right)  ^{2}\gg p^{2}$ and by
Eq.(\ref{yi}), we can do the expansion of Eq.(\ref{rev}) to get%
\begin{equation}
\frac{-K_{1}^{25}K_{2}^{25}(1+\frac{p^{2}}{2(K_{1}^{25})^{2}})(1+\frac{p^{2}%
}{2(K_{2}^{25})^{2}})+p^{2}}{K_{1}^{25}K_{2}^{25}}=-1,
\end{equation}
which in turn, to the first order of the expansion, gives%
\begin{equation}
-K_{1}^{25}K_{2}^{25}(1+\frac{p^{2}}{2(K_{1}^{25})^{2}}+\frac{p^{2}}%
{2(K_{2}^{25})^{2}})+p^{2}=-K_{1}^{25}K_{2}^{25}.
\end{equation}
A simple calculation then gives%
\begin{equation}
(\lambda_{1}+\lambda_{2})^{2}=0 \label{lamda}%
\end{equation}
where signs of $\lambda_{1}=\frac{p}{K_{1}^{25}}$ and $\lambda_{2}=-\frac
{p}{K_{2}^{25}}$ are chosen to be the same. It can be seen now that the
kinematic regime in Eq.(\ref{newpower}) does solve Eq.(\ref{lamda}). In
conclusion, there is a\textit{ }$\phi=$ $finite$\textit{ }regime with UV
power-law behavior for the high-energy compactified open string
scatterings.\textit{ }This new phenomenon never happens in the 26D string
scatterings. The linear relations break down as expected in this regime.

\bigskip%

\setcounter{equation}{0}
\renewcommand{\theequation}{\arabic{section}.\arabic{equation}}%

\section{Regge Scatterings}

We now begin to consider the compactified Regge string scatterings. It is
important at this point to note that in the high-energy, $t_{25}=finite$
approximation, all $\hat{M}_{i}^{2}$ can be neglected and we have%
\begin{equation}
\cos\phi\simeq1+\frac{t_{25}}{2s_{25}}\text{, }p\sin\phi\simeq\sqrt{-{t}_{25}}
\label{change}%
\end{equation}
where we have used Eq.(\ref{new1}) to Eq.(\ref{new4}) to do the calculation.
It is easy to see that high-energy fixed $t_{25}=finite$ (instead of fixed
$t$) approximation corresponds to the \textit{small angle} $\phi$ or Regge
regime (RR). In the high-energy limit, $p^{2}=q^{2}=s_{25}/4.$ In this paper,
we are going to consider two different Regge regimes (RR) corresponding to
fixed winding modes $(K_{i}^{25})^{2}\ll p^{2}$ and highly winding modes
$(K_{i}^{25})^{2}\simeq p^{2}$ respectively.

\subsection{Fixed Winding Modes}

We first consider the following RR%
\begin{equation}
t_{25}=finite,(K_{i}^{25})^{2}\ll p^{2}\gg N. \label{RR1}%
\end{equation}
In this regime%
\begin{equation}
t_{25}\simeq t+(K_{2}^{25}-K_{3}^{25})^{2}. \label{t25}%
\end{equation}
A class of high-energy vertex at fixed mass level $N=\sum_{n,m}np_{n}+mq_{m}$
are \cite{bosonic,RRsusy}%
\begin{equation}
\left\vert p_{n},q_{m},i,l\right\rangle =\prod_{l>0}(\alpha_{-n}^{T})^{p_{n}%
}\prod_{m>0}(\alpha_{-m}^{L})^{q_{m}}\left\vert k_{2},l_{2},i,l\right\rangle .
\end{equation}
The conformal invariant property of the above vertex was discussed in
\cite{RRsusy}. Note that states containing operators $\left(  \alpha_{-n}%
^{25}\right)  $ are of sub-leading order in energy and are neglected. For
simplicity, we will only consider the states%
\begin{equation}
\left\vert N,2m,r,i,l\right\rangle =\left(  \alpha_{-1}^{T}\right)
^{N-2m-2r}\left(  \alpha_{-1}^{L}\right)  ^{2m}\left(  \alpha_{-2}^{L}\right)
^{r}\left\vert k_{2},l_{2},i,l\right\rangle . \label{vertex}%
\end{equation}
at mass level $\hat{M}_{2}^{2}=2\left(  N-1\right)  $ scattered with three
"tachyon" states (with $\hat{M}_{1}^{2}=\hat{M}_{3}^{2}=\hat{M}_{4}^{2}=-2$).
Eq.(\ref{vertex}) is the most general high-energy vertex in the fixed angle
regime. The vertex considered previously at fixed angle in Eq.(\ref{Nr})
corresponds to $m=0$ only and thus was not completed$.$ The relevant
kinematics can be calculated to be%
\begin{equation}
e^{P}\cdot k_{1}\simeq-\frac{s_{25}}{2M_{2}},\text{ \ }e^{P}\cdot k_{3}%
\simeq-\frac{t_{25}-M_{2}^{2}-M_{3}^{2}}{2M_{2}}=-\frac{\tilde{t}_{25}}%
{2M_{2}};
\end{equation}%
\begin{equation}
e^{L}\cdot k_{1}\simeq-\frac{s_{25}}{2M_{2}},\text{ \ }e^{L}\cdot k_{3}%
\simeq-\frac{t_{25}+M_{2}^{2}-M_{3}^{2}}{2M_{2}}=-\frac{\tilde{t}_{25}%
^{\prime}}{2M_{2}};
\end{equation}
and%
\begin{equation}
e^{T}\cdot k_{1}=0\text{, \ \ }e^{T}\cdot k_{3}\simeq-\sqrt{-{t}_{25}}.
\end{equation}

We are now ready to calculate the Regge scattering amplitudes. Note that
$e^{P}\neq e^{L}$ in the RR \cite{bosonic,bosonic2,RRsusy}. We will calculate
$e^{L}$ amplitudes in ths paper. The corresponging $e^{P}$ amplitudes can be
similarly calculated. The $s-t$ channel of the compactified Regge string
scattering amlpitudes in the regime Eq.(\ref{RR1}) can be calculated to be (We
will ignore the trace factor due to Chan-Paton in the scattering amplitude
calculation . This does not affect our final results in this paper)%
\begin{align}
A^{(N,2m,r,i,l)}  &  =\int_{0}^{1}dx\,x^{k_{1}\cdot k_{2}}(1-x)^{k_{2}\cdot
k_{3}}\left[  \frac{e^{T}\cdot k_{3}}{1-x}\right]  ^{N-2m-2r}\nonumber\\
\cdot &  \left[  \frac{e^{L}\cdot k_{1}}{-x}+\frac{e^{L}\cdot k_{3}}%
{1-x}\right]  ^{2m}\left[  \frac{e^{L}\cdot k_{1}}{x^{2}}+\frac{e^{L}\cdot
k_{3}}{(1-x)^{2}}\right]  ^{r}\nonumber\\
&  \simeq(\sqrt{-{t}_{25}})^{N-2m-2r}\left(  \frac{\tilde{t}_{25}^{\prime}%
}{2M_{2}}\right)  ^{r}\int_{0}^{1}dx\,x^{k_{1}\cdot k_{2}}(1-x)^{k_{2}\cdot
k_{3}-N+2m}\nonumber\\
\cdot &  \sum_{j=0}^{2m}{\binom{2m}{j}}\left(  \frac{s_{25}}{2M_{2}x}\right)
^{j}\left(  \frac{-\tilde{t}_{25}^{\prime}}{2M_{2}(1-x)}\right)
^{2m-j}\nonumber\\
&  =(\sqrt{-{t}_{25}})^{N-2m-2r}\left(  \frac{\tilde{t}_{25}^{\prime}}{2M_{2}%
}\right)  ^{r}\left(  \frac{\tilde{t}_{25}^{\prime}}{2M_{2}}\right)
^{2m}\nonumber\\
\cdot &  \sum_{j=0}^{2m}{\binom{2m}{j}}(-1)^{j}\left(  \frac{s_{25}}{\tilde
{t}_{25}^{\prime}}\right)  ^{j}B\left(  k_{1}\cdot k_{2}-j+1,k_{2}\cdot
k_{3}-N+j+1\right)  . \label{7}%
\end{align}
Note that the term $\frac{e^{L}\cdot k_{1}}{x^{2}}$ in the bracket is
subleading in energy and can be neglected. In the high-energy limit, the beta
function in Eq.(\ref{7}) can be approximated by%
\begin{equation}
B\left(  k_{1}\cdot k_{2}-j+1,k_{2}\cdot k_{3}-N+j+1\right)  \simeq B\left(
-1-\frac{1}{2}s,-1-\frac{t}{2}\right)  \left(  -\frac{s}{2}\right)
^{-j}\left(  -1-\frac{t}{2}\right)  _{j}. \label{8}%
\end{equation}
Finally, the leading order amplitude in the RR can be written as%
\begin{align}
A^{(N,2m,r,i,l)}  &  =B\left(  -1-\frac{s}{2},-1-\frac{t}{2}\right)
\sqrt{-t_{25}}^{N-2m-2r}\left(  \frac{1}{2M_{2}}\right)  ^{2m+r}\nonumber\\
&  2^{2m}(\tilde{t}_{25}^{\prime})^{r}U\left(  -2m\,,\,\frac{t}{2}%
+2-2m\,,\,\frac{\tilde{t}_{25}^{\prime}}{2}\right)  , \label{9}%
\end{align}
which is UV power-law behaved as $t=finite$ in the beta function by
Eq.(\ref{t25}). $U$ in Eq.(\ref{9}) is the Kummer function of the second kind
and is defined to be%
\begin{equation}
U(a,c,x)=\frac{\pi}{\sin\pi c}\left[  \frac{M(a,c,x)}{(a-c)!(c-1)!}%
-\frac{x^{1-c}M(a+1-c,2-c,x)}{(a-1)!(1-c)!}\right]  \text{ \ }(c\neq2,3,4...)
\label{10}%
\end{equation}
where $M(a,c,x)=\sum_{j=0}^{\infty}\frac{(a)_{j}}{(c)_{j}}\frac{x^{j}}{j!}$ is
the Kummer function of the first kind. $U$ and $M$ are the two solutions of
the Kummer equation%
\begin{equation}
xy^{^{\prime\prime}}(x)+(c-x)y^{\prime}(x)-ay(x)=0. \label{11}%
\end{equation}
It is crucial to note that, in our case of Eq.(\ref{9}), $c=\frac{t}{2}+2-2m$
and is not a constant as in the usual definition, so $U$ in Eq.(\ref{9})
is\textit{ not} a solution of the Kummer equation.

It is important to note that there is no linear relation among high-energy
string scattering amplitudes of different string states for each fixed mass
level in the RR as can be seen from Eq.(\ref{9}). This is very different from
the result in the GR in Eq.(\ref{2}). In other words, the ratios
$A^{(N,2m,r,i,l)}/A^{(N,0,0,i,l)}$ are $\tilde{t}_{25}^{\prime}$-dependent
functions. In particular, we can extract the coefficients of the highest power
of $\tilde{t}_{25}^{\prime}$ in $A^{(N,2m,r,i,l)}/A^{(N,0,0,i,l)}$. We can use
the identity of the Kummer function%
\begin{align}
&  2^{2m}(\tilde{t}_{25}^{\prime})^{-2m}\ U\left(  -2m,\frac{t}{2}%
+2-2m,\frac{\tilde{t}_{25}^{\prime}}{2}\right) \nonumber\\
&  =\,_{2}F_{0}\left(  -2m,-1-\frac{t}{2},-\frac{2}{\tilde{t}_{25}^{\prime}%
}\right) \nonumber\\
&  \equiv\sum_{j=0}^{2m}\left(  -2m\right)  _{j}\left(  -1-\frac{t}{2}\right)
_{j}\frac{\left(  -\frac{2}{\tilde{t}_{25}^{\prime}}\right)  ^{j}}%
{j!}\nonumber\\
&  =\sum_{j=0}^{2m}{\binom{2m}{j}}\left(  -L-\frac{\tilde{t}_{25}^{\prime}}%
{2}\right)  _{j}\left(  \frac{2}{\tilde{t}_{25}^{\prime}}\right)  ^{j}
\label{12}%
\end{align}
to get
\begin{align}
\frac{A^{(N,2m,r,i,l)}}{A^{(N,0,0,i,l)}}  &  =(-1)^{m}\left(  -\frac{1}%
{2M_{2}}\right)  ^{2m+r}(\tilde{t}_{25}^{\prime}-M_{2}^{2}+M_{3}^{2}%
)^{-m-r}(\tilde{t}_{25}^{\prime})^{2m+r}\nonumber\\
&  \cdot\sum_{j=0}^{2m}(-2m)_{j}\left(  -L-\frac{\tilde{t}_{25}^{\prime}}%
{2}\right)  _{j}\frac{(-2/\tilde{t}_{25}^{\prime})^{j}}{j!}+\mathit{O}\left\{
\left(  \frac{1}{\tilde{t}_{25}^{\prime}}\right)  ^{m+1}\right\}  \label{13}%
\end{align}
where%
\begin{equation}
L=1-N-(K_{2}^{25})^{2}+K_{2}^{25}K_{3}^{25}. \label{L}%
\end{equation}
If the leading order coefficients in Eq.(\ref{13}) extracted from the high
energy string scattering amplitudes in the RR are to be identified with the
complete ratios in Eq.(\ref{2}) calculated previously among high energy string
scattering amplitudes in the GR \cite{bosonic,bosonic2}%
\begin{equation}
\lim_{\tilde{t}_{25}^{\prime}\rightarrow\infty}\frac{A^{(N,2m,r,i,l)}%
}{A^{(N,0,0,i,l)}}=\left(  -\frac{1}{M_{2}}\right)  ^{2m+r}\left(  \frac{1}%
{2}\right)  ^{m+r}(2m-1)!!=\frac{T^{(N,2m,r,i,l)}}{T^{(N,0,0,i,l)}},
\label{Ratios2}%
\end{equation}
we need the following identity
\begin{align}
&  \sum_{j=0}^{2m}(-2m)_{j}\left(  -L-\frac{\tilde{t}_{25}^{\prime}}%
{2}\right)  _{j}\frac{(-2/\tilde{t}_{25}^{\prime})^{j}}{j!}\nonumber\\
&  =0(-\tilde{t}_{25}^{\prime})^{0}+0(-\tilde{t}_{25}^{\prime})^{-1}%
+...+0(-\tilde{t}_{25}^{\prime})^{-m+1}+\frac{(2m)!}{m!}(-\tilde{t}%
_{25}^{\prime})^{-m}+\mathit{O}\left\{  \left(  \frac{1}{\tilde{t}%
_{25}^{\prime}}\right)  ^{m+1}\right\}  . \label{14}%
\end{align}
Note that the ratios calculated previously at fixed angle in Eq.(\ref{linear})
corresponds to $m=0$ only in Eq.(\ref{Ratios2}) and thus was not completed.
The ratios in Eq.(\ref{Ratios2}) calculated in this paper by the indirect
method through the RR amplitudes are the most general ones. The coefficient of
the term $\mathit{O}\left\{  \left(  1/\tilde{t}_{25}^{\prime}\right)
^{m+1}\right\}  $ in Eq.(\ref{14}) is irrelevant for our discussion. The proof
of Eq.(\ref{14}) turns out to be nontrivial. The standard approach by using
integral representation of the Kummer function seems not applicable here.
Presumably, the difficulty of the rigorous proof of Eq.(\ref{14}) is
associated with the nonconstant $c$ mentioned previously.

Mathematically, the complete proof of Eq.(\ref{14}) for \textit{arbitrary real
values} $L$ was recently worked out in \cite{LYAM} by using an identity of
signless Stirling number of the first kind in combinatorial theory. The proof
of the identity for $L=0,1,$ was previously given in
\cite{bosonic,bosonic2,RRsusy} based on a set of identities of signed Stirling
number of the first kind \cite{MK}. It is interesting to see that, physically,
the identities for arbitrary real values $L$ can only be realized in
high-energy compactified string scatterings considered in this paper. This is
due to the dependence of the value $L$ on winding momenta $K_{i}^{25}.$ All
other high-energy string scattering amplitudes calculated previously
\cite{bosonic,bosonic2,RRsusy} correspond to integer value of $L$ only.

\subsection{Highly Winding Modes}

In this subsection, we consider the more interesting RR%
\begin{equation}
t_{25}=finite,(K_{i}^{25})^{2}\simeq p^{2}\gg N. \label{RR2}%
\end{equation}
In this regime, Eq.(\ref{t}) and Eq.(\ref{new2}) imply%
\begin{equation}
t_{25}\simeq t-\sqrt{p^{2}+\hat{M}_{2}^{2}}\cdot\sqrt{q^{2}+\hat{M}_{3}^{2}%
}+\sqrt{p^{2}+M_{2}^{2}}\cdot\sqrt{q^{2}+M_{3}^{2}}-K_{2}^{25}K_{3}^{25}.
\end{equation}
It's easy to see that in general\footnote{For some regime, $t$ can be finite.
For example, for $K_{2}^{25}=K_{3}^{25}\simeq p^{2}\gg N,t\simeq
t_{25}=finite$ by Eq.(4.20).} $t$ is as large as $p^{2}$ in this regime. The
most general high-energy vertex at each fixed mass level $N$ are%
\begin{equation}
\left\vert N,2m,r,i,l\right\rangle =\left(  \alpha_{-1}^{T}\right)
^{N-2m-2r}\left(  \alpha_{-1}^{L}\right)  ^{2m}\left(  \alpha_{-2}^{L}\right)
^{r}\left\vert k_{2},l_{2},i,l\right\rangle .
\end{equation}
Note that states containing operators $\left(  \alpha_{-n}^{25}\right)  $ are
again of sub-leading order in energy. For simplicity, we will only consider
the states%
\begin{equation}
\left\vert N,0,r,i,l\right\rangle =\left(  \alpha_{-1}^{T}\right)
^{N-2r}\left(  \alpha_{-2}^{L}\right)  ^{r}\left\vert k_{2},l_{2}%
,i,l\right\rangle
\end{equation}
at mass level $\hat{M}_{2}^{2}=2\left(  N-1\right)  $ scattered with three
"tachyon" states (with $\hat{M}_{1}^{2}=\hat{M}_{3}^{2}=\hat{M}_{4}^{2}=-2$).
The $s-t$ channel of the high-energy scattering amlpitude can be calculated to
be
\begin{align}
A^{(N,0,r,i,l)}  &  =\int d^{4}x\cdot\prod\limits_{i<j}\left(  x_{i}%
-x_{j}\right)  ^{k_{i}\cdot k_{j}}\nonumber\\
&  \cdot\left[  \frac{ie^{T}\cdot k_{1}}{x_{1}-x_{2}}+\frac{ie^{T}\cdot k_{3}%
}{x_{3}-x_{2}}+\frac{ie^{T}\cdot k_{4}}{x_{4}-x_{2}}\right]  ^{N-2r}%
\cdot\left[  \frac{e^{L}\cdot k_{1}}{\left(  x_{1}-x_{2}\right)  ^{2}}%
+\frac{e^{L}\cdot k_{3}}{\left(  x_{3}-x_{2}\right)  ^{2}}+\frac{e^{L}\cdot
k_{4}}{\left(  x_{4}-x_{2}\right)  ^{2}}\right]  ^{r}.
\end{align}
After fixing the $SL(2,R)$ gauge and using the kinematic relations
Eq.(\ref{KR1}) to Eq.(\ref{KR6}) and Eq.(\ref{change}) derived previously, we
have%
\begin{align}
A^{(N,0,r,i,l)}  &  =\left(  -i\sqrt{-t_{25}}\right)  ^{N}\left(  \frac
{p\sqrt{q^{2}+M_{3}^{2}}-q\sqrt{p^{2}+M_{2}^{2}}}{M_{2}t_{25}}\right)
^{r}\nonumber\\
&  \cdot\int_{0}^{1}dx\cdot x^{k_{1}\cdot k_{2}}\left(  1-x\right)
^{k_{2}\cdot k_{3}-N+2r}\nonumber\\
&  \cdot\left[  \frac{p\sqrt{p^{2}+M_{1}^{2}}+p\sqrt{p^{2}+M_{2}^{2}}}{\left(
p\sqrt{q^{2}+M_{3}^{2}}-q\sqrt{p^{2}+M_{2}^{2}}\right)  x^{2}}-\frac
{1}{\left(  1-x\right)  ^{2}}\right]  ^{r}\nonumber\\
&  =\left(  -i\sqrt{-t_{25}}\right)  ^{N}\left(  \frac{p\sqrt{q^{2}+M_{3}^{2}%
}-q\sqrt{p^{2}+M_{2}^{2}}}{M_{2}t_{25}}\right)  ^{r}\nonumber\\
&  \cdot\sum_{j=0}^{r}\binom{r}{j}\left[  -\frac{p\sqrt{p^{2}+M_{1}^{2}%
}+p\sqrt{p^{2}+M_{2}^{2}}}{p\sqrt{q^{2}+M_{3}^{2}}-q\sqrt{p^{2}+M_{2}^{2}}%
}\right]  ^{j}\cdot\int_{0}^{1}dx\cdot x^{k_{1}\cdot k_{2}-2j}\left(
1-x\right)  ^{k_{2}\cdot k_{3}-N+2j}\nonumber\\
&  =\left(  -i\sqrt{-t_{25}}\right)  ^{N}\left(  \frac{p\sqrt{q^{2}+M_{3}^{2}%
}-q\sqrt{p^{2}+M_{2}^{2}}}{M_{2}t_{25}}\right)  ^{r}\nonumber\\
&  \cdot\sum_{j=0}^{r}\binom{r}{j}\left[  -\frac{p\sqrt{p^{2}+M_{1}^{2}%
}+p\sqrt{p^{2}+M_{2}^{2}}}{p\sqrt{q^{2}+M_{3}^{2}}-q\sqrt{p^{2}+M_{2}^{2}}%
}\right]  ^{j}\nonumber\\
&  \cdot B\left(  -\frac{1}{2}s+N-2j-1,-\frac{1}{2}t+2j-1\right)
\end{align}
where $B(u,v)$ is the Euler beta function. We can do the high-energy
approximation of the gamma function $\Gamma\left(  x\right)  $ and end up
with
\begin{align}
A^{(N,0,r,i,l)}  &  =\left(  -i\sqrt{-t_{25}}\right)  ^{N}\left(  \frac
{p\sqrt{q^{2}+M_{3}^{2}}-q\sqrt{p^{2}+M_{2}^{2}}}{M_{2}t_{25}}\right)
^{r}\nonumber\\
&  \cdot\sum_{j=0}^{r}\binom{r}{j}\left[  -\frac{p\sqrt{p^{2}+M_{1}^{2}%
}+p\sqrt{p^{2}+M_{2}^{2}}}{p\sqrt{q^{2}+M_{3}^{2}}-q\sqrt{p^{2}+M_{2}^{2}}%
}\right]  ^{j}\nonumber\\
&  \cdot\frac{\Gamma\left(  -1-\frac{1}{2}s+N-2j\right)  \Gamma\left(
-1-\frac{1}{2}t+2j\right)  }{\Gamma\left(  2+\frac{1}{2}u\right)  }\nonumber\\
&  \simeq\left(  -i\sqrt{-t_{25}}\right)  ^{N}\left(  \frac{p\sqrt{q^{2}%
+M_{3}^{2}}-q\sqrt{p^{2}+M_{2}^{2}}}{M_{2}t_{25}}\right)  ^{r}\nonumber\\
&  \cdot\sum_{j=0}^{r}\binom{r}{j}\left[  -\frac{p\sqrt{p^{2}+M_{1}^{2}%
}+p\sqrt{p^{2}+M_{2}^{2}}}{p\sqrt{q^{2}+M_{3}^{2}}-q\sqrt{p^{2}+M_{2}^{2}}%
}\right]  ^{j}\nonumber\\
&  \cdot B\left(  -1-\dfrac{1}{2}s,-1-\frac{1}{2}t\right)  \left(  -1-\frac
{1}{2}t\right)  _{2j}\left(  -1-\dfrac{1}{2}s\right)  ^{N-2j}\left(
2+\dfrac{1}{2}u\right)  ^{-N}\nonumber\\
&  =\left(  i\frac{\sqrt{-t_{25}}}{(\frac{u}{s})}\right)  ^{N}\left(
\frac{p\sqrt{q^{2}+M_{3}^{2}}-q\sqrt{p^{2}+M_{2}^{2}}}{M_{2}t_{25}}\right)
^{r}\cdot B\left(  -1-\dfrac{1}{2}s,-1-\frac{1}{2}t\right) \nonumber\\
&  \cdot\sum_{j=0}^{r}\binom{r}{j}\left[  -\frac{p\sqrt{p^{2}+M_{1}^{2}%
}+p\sqrt{p^{2}+M_{2}^{2}}}{p\sqrt{q^{2}+M_{3}^{2}}-q\sqrt{p^{2}+M_{2}^{2}}%
}\dfrac{4}{s^{2}}\right]  ^{j}\left(  -1-\frac{1}{2}t\right)  _{2j}
\label{general}%
\end{align}
Finally, since $t$ is as large as $p^{2}$ in the regime Eq.(\ref{KR1}), we can
easily do the summation and end up with%

\begin{align}
A^{(N,0,r,i,l)}  &  =\left(  i\frac{\sqrt{-t_{25}}}{(\frac{u}{s})}\right)
^{N}\left(  -\frac{1}{M_{2}}\right)  ^{r}\cdot B\left(  -1-\dfrac{1}%
{2}s,-1-\frac{1}{2}t\right) \nonumber\\
&  \left[  -\frac{p\sqrt{q^{2}+M_{3}^{2}}-q\sqrt{p^{2}+M_{2}^{2}}}{t_{25}%
}+\frac{p\sqrt{p^{2}+M_{1}^{2}}+p\sqrt{p^{2}+M_{2}^{2}}}{t_{25}}\left(
\dfrac{t}{s}\right)  ^{2}\right]  ^{r} \label{end}%
\end{align}
where $(\frac{t}{s})$ and $(\frac{u}{s})$ are fixed numbers. Since $t$ is as
large as $s$ in this regime, the beta function in Eq.(\ref{end}) implies that
the UV behavior of the amplitude is exponential fall-off. On the other hand,
it is clearly that there is no linear relation in this regime. In conclusion,
we have discovered a $\phi\simeq0$ regime with UV exponential fall-off
behavior for the high-energy compactified open string scatterings. This new
phenomenon never happens in the 26D string scatterings.%

\setcounter{equation}{0}
\renewcommand{\theequation}{\arabic{section}.\arabic{equation}}%

\section{Conclusion}

In this paper, we have mainly achieved three new results for high-energy
string scattering amplitudes. First, we calculate massive string scattering
amplitudes of compactified open string in the Regge regime. We can then
extract the \textit{complete} infinite ratios among high-energy amplitudes of
different string states in the fixed angle regime from these Regge string
scattering amplitudes. The complete ratios calculated by this indirect method
include and extend the subset of ratios calculated previously
\cite{Compact,Compact2} by the more difficult direct fixed angle calculation.

Secondly, by studying the high-energy string scattering for the
\textit{compactified} open string, we discover in this paper a realization of
arbitrary real values $L$ in the identity Eq.(\ref{14}) which was proposed
recently to link fixed angle and Regge string scattering amplitudes. All other
high-energy string scatterings calculated previously \cite{bosonic,RRsusy,LMY}
correspond to integer value of $L$ only. Physically , The parameter $L$ is
related to the mass level of an excited string state and can take non-integer
values for Kaluza-Klein modes. Mathematically, the identity in Eq.(\ref{14})
was explicitly proved recently for arbitrary real values $L$ in \cite{LYAM} by
using the signless Stirling number in combinatorial theory.

Finally, we discover a kinematic regime which shows the unusual exponential
fall-off behavior in the small angle scattering. This is in complementary with
a fixed angle regime discovered previously \cite{Compact2}, which shows the
unusual power-law behavior in the compactified string scatterings.

\section{Acknowledgments}

This work is supported in part by the National Science Council, 50 Billions
Project of MOE and National Center for Theoretical Science, Taiwan, R.O.C. We
would like to thank the hospitality of KITPC where part of this work was
completed during our visits in the summer of 2010. J.C. Lee and Yi Yang would
like to thank discussions on this subject with Prof. C.I. Tan of Brown
University, Prof. B. Feng of Zhejiang University and Dr.Y. Mitsuka. S. He
would like to thank Prof. Mei Huang and Prof. Sang Jin Sin's warm support. He
is grateful to APCTP and CQUeST in Korea for their hospitalities at various
stages of this work.

\end{document}